\documentstyle[prd,aps,amsfonts]{revtex}
\begin{document}
\draft
\title{Properties of Intersecting p-branes in Various Dimensions}
\author{I.~Ya.~Aref'eva\thanks { e-mail: arefeva@arevol.mian.su}}
\address{Steklov Mathematical Institute,
         Gubkin 8, GSP-1, 117966, Moscow, Russia}
\author{M.~G.~Ivanov\thanks{e-mail: mgi@landau.ac.ru}}
\address{Department of General and Applied Physics,
      Moscow Institute of Physics and Technology,\\
      Instituski per., 9, Dolgoprudnyi, Moscow region, Russia}
\author{O.~A.~Rytchkov\thanks { e-mail: rytchkov@grg1.phys.msu.su}}
\address{Department of Theoretical Physics,
         Moscow State University,\\ Moscow 119 899, Russia}
\date {\today}
\maketitle
\begin {abstract}
 General properties of intersecting extremal p-brane
 solutions of gravity coupled with dilatons and several
 different d-form fields in arbitrary
 space-time dimensions are considered. It is shown
 that heiristically expected properties of the intersecting
 p-branes follow from the explicit formulae for solutions.
 In particular, harmonic superposition and $S$-duality
 hold for all p-brane solutions. Generalized $T$-duality
 takes place under additional restrictions on the initial
 theory parameters.
\end {abstract}

\section{Introduction}
 There has been recently considerable progress in the
 study of classical p-brane solutions
 (for review see \cite{Stelle} and references therein)
 of higher dimensional gravity coupled with matter.
 p-brane solutions in 10 and 11 dimensions play a key role
 for probing the duality conjectures \cite{HT,W,PT,JHS,Sen}
 which relate five known superstrings and M-theory.
 A study of p-brane intersections
 is a subject of growing interest because when enough branes intersect
 one gets, as a rule, solutions
 with a regular horizon. The  microscopic
 interpretation of the Bekenstein-Hawking entropy within
 string theory \cite{Vafa} has also stimulated  investigations of the
 intersecting (composite) p-brane solutions. Several composite
 p-brane solutions in $D=10$ and $D=11$
 have been obtained
\cite{TS,CM,TP1,Ts1,Pap,klT,BB,Gn,berg,Ts2,BO,Hor,Lust,Em,kl,Berg,PT2,Ts212}.

 Heuristic scheme of constructing of p-brane intersections
 was based on string theory representation of the branes,
 duality and supersymmetry. This scheme involves the
 harmonic function superposition rule for the intersecting p-branes.
 This rule was formulated \cite{Ts1} in $D=11$ and $D=10$.
 Using T-duality and the
 supersymmetry requirements intersections of p-branes
 (more exactly M-branes and D-branes) have been recently
 classified in \cite{BergRoo}. In \cite{AEH} it was shown,
 that the intersection rules applying to type II string
 theories and M-theory are consistent with the picture
 that open brane can have boundaries on some other branes.

 Other approach to the problem was elaborated in the papers
 \cite{AV,AVV,IN,IvMel,ArRy,AVVV,AIV,Ohta,Angle}.
 In these papers p-brane intersection rules were found from
 the equations of motion. The aim of this paper is
 to summarize these results, and to show that heiristically
 expected properties of the intersecting p-branes follow from
 the explicit formulae for solutions.

The paper is organized as follows. In Section 2 we remind single
p-brane solutions and introduce graphic representations for them.
In Section 3 the main steps of finding p-brane solutions are sketched
and general composite p-brane solutions are presented. In Section 4
we collect the formulae for the entropy and ADM mass. In the section
5 we discuss $S$-duality, which is a specific property of our solutions.
Harmonic function rule is generalized on arbitrary space-time dimensions
in Section 6. $T$-duality transformations are considered in Section 7.
In Section 8 we modify our results for the case of an arbitrary
space--time signature. In Section 9 we analyze supersymmetry in the
special case of 11D supergravity.

\section{Single p-brane solutions}
 Let us consider the theory with the following action
 \begin{equation}\label{1}
  I=\frac{1}{2\kappa ^{2}}\int d^{D}X\sqrt{-g}
    \left(
      R- \frac{1}{2}(\nabla \phi)^{2}-\frac{e^{-\alpha \phi}}
      {2(d+1)!}F^{2}_{d+1}\right),
 \end{equation}
 where $F_{d+1}$ is a $d+1$ differential form,
 $F_{d+1}=dA_d$, $\phi$ is a dilaton.

 This action admits the  elementary p-brane solution \cite{DGHR}
 \begin{equation}\label{el1}
   ds^2=H^\tau(x)\sum_{\mu,\nu=0}^{d-1}
        \eta_{\mu\nu}dy^{\mu}dy^{\nu}+
        H^\rho(x)\sum_{\gamma=1}^{D-d}
        dx^{\gamma}dx^{\gamma},\
 \end{equation}
 \begin{equation}\label{gf}
   A_d=2\Delta^{-1/2}
       H^{-1}(x)dy^0\wedge\dots\wedge dy^{d-1},
 \end{equation}
 where $H(x)$ is a harmonic function.
 Solution (\ref{el1}), (\ref{gf}) generalizes the
 well-known Majumdar-Papapetrou solutions \cite{MP}.
 The exponents are defined by the parameters of the
 theory
 \begin{equation}
   \tau =-{4(D-2-d)\over \Delta(D-2)},\quad
   \rho ={4d\over \Delta(D-2)},
 \end{equation}
 where we use the standard notation
 \begin{equation}\label{Dlt}
   \Delta=\alpha^2+{2d(D-2-d)\over D-2}.
 \end{equation}

 In the p-brane terminology \cite{Stelle} the solution
 (\ref{el1})--(\ref{Dlt}) describes an electrically
 charged p-brane, where $y$-coordinates correspond to
 the worldvolume directions and $x$-coordinates to directions
 transverse to the brane.  It is convenient to represent every
 $y^k$-coordinate by ``$\times$'' and every $x$-coordinate by
 ``--''~\cite{BergRoo}. One has the following representation
 of the metric
 \begin{equation}
   ds^2=\underbrace{\times\;\times\cdots\;\times}_d
        \underbrace{\;-\;-\cdots -}_{D-d}.
 \end{equation}
 It is also convenient to present the gauge
 field (\ref{gf}) as a row with d circles "$\circ$"
\vspace{10pt}
 \begin{center}
   \begin{tabular}{|cccc|ccc|}
     \hline
     $\circ$&$\circ$&$\cdots$&$\circ$&&&\\
     \hline
   \end{tabular}

   \vspace{5pt}
 Fig.1
  \end{center}
  \vspace{5pt}
 where the circles correspond to indices of the non-zero
component.

 The action (\ref{1}) admits also a solitonic p-brane solution
 \cite{Callan-1991}
 \begin{equation}\label{sol1}
   ds^2=U^{-\rho}(x)\sum_{\mu,\nu=0}^{D-d-3}
        \eta_{\mu\nu}dy^{\mu}dy^{\nu}+
        U^{-\tau}(x)\sum_{\gamma=1}^{d+2}
        dx^{\gamma}dx^{\gamma},\
 \end{equation}
 \begin{equation}\label{solF}
   F=2\Delta^{-1/2} \ast dU,
 \end{equation}
 where $\ast$ is a Hodge dual on
 ${\Bbb{R}}^{d+2}$, $U(x)$ is a harmonic function.
 Using the p-brane interpretation one says that
 the solution (\ref{sol1}), (\ref{solF}) describes the
 magnetically charged p-brane lying in the $y$-directions.
 One can present the metric
 in the following way
 \begin{equation}
   ds^2=\underbrace{\ast\;\ast\cdots\;\ast}_{D-d-2}
        \underbrace{\;-\;-\cdots -}_{d+2},
 \end{equation}
 where ``$\ast$'' denote the worldvolume directions.
 Note that we use the symbol ``$\times$''
 for the electric p-brane and the
 symbol ``$\ast$'' for the magnetic one.

 For the gauge field we have a different picture with $d+2$
 circles "$\bullet$"
  \noindent
 \vspace{10pt}
 \begin{center}
   \begin{tabular}{|ccc|cccc|}
     \hline
     &&&$\bullet$&$\bullet$&$\cdots$&$\bullet$\\
     \hline
   \end{tabular}

  \vspace{5pt}
Fig.2
 \end{center}
 \vspace{5pt}
 where by ``$\bullet$'' we denote subspace
 on which a Hodge dual acts. For the more complicated
 ansatzes we have pictures with more then one rows.
 For example,
  \noindent
 \vspace{10pt}
 \begin{center}
   \begin{tabular}{|c|c|c|c|c|ccc|c|c|c|}
     \hline
     $\circ$&$\circ$&$\circ$&&&&&&&&\\
     \hline
     $\circ$&&&$\circ$&$\circ$&&&&&&\\
     \hline
     &$\bullet$&&$\bullet$&&&&&$\bullet$&$\bullet$&$\bullet$\\
     \hline
     &&$\bullet$&&$\bullet$&&&&$\bullet$&$\bullet$&$\bullet$\\
     \hline
   \end{tabular}

   \vspace{5pt}
 Fig.3
 \end{center}
 \vspace{5pt}
 We will call such pictures for the antisymmetric field as
 incidence tables. In the p-brane terminology the
 corresponding metric could be presented as
 \vspace{5pt}
 \begin{center}
   \begin{equation}\label{brtab2}
     ds^2=
     \left\{
       \begin{tabular}{c|c|c|c|c|ccc|c|c|c|}
         $\times$&$\times$&$\times$&--&--&--&--&--&--&--&--\\
         $\times$&--&--&$\times$&$\times$&--&--&--&--&--&--\\
         $\ast$&--&$\ast$&--&$\ast$&$\ast$&$\ast$&$\ast$&--&--&--\\
         $\ast$&$\ast$&--&$\ast$&--&$\ast$&$\ast$&$\ast$&--&--&--\\
       \end{tabular}
     \right..
   \end{equation}

   \vspace{5pt}
 \end{center}
 \vspace{5pt}

 We will call such pictures for the metric as
 {\it brane incidence tables}.
 The intersections of the p-branes are not arbitrary
 and governed by characteristic equations which are
 considered below.

\section{Composite p-brane solutions}

 The method \cite{AV,AVV,IN,ArRy} of finding the
 intersecting p-brane solutions
 involves direct solving the equations of motion for the theory
 (\ref{1}). The method consists of the following steps.
 \par {\bf Step 1}.
 We assume the metric  in the special form (see (\ref{3})).
 In order to simplify the Ricci tensor we also assume that the
 form of the metric satisfies the Fock--De~Donder gauge condition
 (see (\ref{gauge}) below).
 \par {\bf Step 2}.
 In order to describe an ansatz for the antisymmetric
 field we introduce the {\it incidence} matrices
 (see (\ref{4e}), (\ref{4m})).
 \par {\bf Step 3}.
 Using assumed form for the antisymmetric field we calculate
 components of the energy-momentum tensor. They could be crucially
 simplified if one assumes so-called ``no-force'' conditions. Their
 origin is simple:  they eliminate exponents from the components of
 the energy-momentum tensor.
 \par {\bf Step 4}. From Maxwell's
 equations and Bianchi identity we conclude that scalar
 functions specifying our ansatzes should be harmonic ones.
 \par {\bf Step 5}.
 We check that Einstein equations and equation of motion for the
 dilaton are fulfilled if the incidence matrices satisfy {\it
 characteristic} equations (see (\ref{7}), (\ref{9}), (\ref{10})).

 Let us briefly demonstrate the realization of this program.
 We assume the metric in the following form
 \begin{equation}\label{3}
   ds^2=e^{2A(x)}\sum_{\mu,\nu=0}^{q-1}
        \eta_{\mu\nu}dy^\mu dy^\nu
        +\sum_{i=q}^{D-s-3}e^{2F_i(x)}dy^i dy^i
        +e^{2B(x)}\sum_{\gamma=D-s-2}^{D-1}
        dx^\gamma dx^\gamma,
 \end{equation}
 where $\eta_{\mu\nu}$ is a flat Minkowski metric, $A$, $B$ and
 $F_{i}$ are functions of $x$. Using the notations
 $F_\mu(x)=A(x)$ and $F_\gamma(x)=B(x)$  the above
 metric may be rewritten as
 \begin{equation}
   ds^2=\sum_{L=0}^{D-1}e^{2F_L(x)}\eta_{KL}dX^K dX^L.
 \end{equation}
 If we assume the Fock--De~Donder gauge condition, which
 for metric (\ref{3}) takes the form
 \begin{equation}\label{gauge}
   \sum_{L=0}^{D-1}F_L-2B=0,
 \end{equation}
 then the Ricci tensor has the following components
 \begin{equation}
   R_{KL}=-\sum_{N=0}^{D-1}\partial_K F_N
                           \partial_L F_N
          +2\partial_K B\partial_L B
          -e^{2F_L-2B}\eta_{KL}\triangle F_L .
 \end{equation}

 For the d-form $A_d$ we consider a class of ansatzes
 corresponding to $E$ electric and $M$ magnetic charges,
 \begin{equation}\label{3'}
  F=\sum _{a}^{E}dA^{(\cal E)}_{a}+
    \sum _{b}^{M} F^{(\cal M)}_{b}.
 \end{equation}
   We will refer to the different terms in (\ref{3'})
  as to the different branches of electric and magnetic fields.

\par
 To describe  electric and magnetic configurations $A^{(\cal E)}_{a}$
 and $F^{(\cal M)}_{b}$ we introduce electric and magnetic
 {\it incidence} matrices
 \begin{eqnarray}\label{4e}
  \Delta &=&(\Delta _{aL}), ~~a=1,\dots,E, ~~L=0,\dots,D-1,\\
  \label{4m}
  \Lambda&=&(\Lambda_{bL}), ~~b=1,\dots,M, ~~L=0,\dots,D-1,
 \end{eqnarray}
  respectively.
 Their rows correspond to independent branches of  the electric
 (magnetic) gauge field and  columns refer to the
 space-time indices.
 The entries of the incidence matrices are equal to 1 or 0.
 Furthermore the electric (magnetic) incidence matrix has an equal
 number of units in each row, and there are no rows which coincide.
 Here we don't consider Euclidean p-branes so we assume
 $\Delta_{a0}=1,\;\Delta_{a\alpha}=0,
 \;\Lambda_{b0}=0\;\Lambda_{b\alpha}=1$ for all $a$ and $b$
 (see section \ref{asig}).
 In a graphic representation  we draw ``$\circ$''
 (for the electric incidence matrix) and ``$\bullet$'' (for the
 magnetic incidence matrix) instead of 1.  Empty space denotes 0 in
 both cases.  Fig.1 and Fig.2 are nothing but a graphic
 representation of the incidence matrices for the single elementary
 and solitonic p-brane solutions. The incidence table Fig.3 is an
 example representing electric and magnetic incidence matrices in a
 more complicated case.

 In the terms of the incidence matrices an electric
 field is assumed to have the form
 \begin{equation}\label{1.1}
   A_a^{({\cal E})}=h_ae^{C_a(x)}
     \bigwedge_{\{L|\Delta_{aL}=1\}}dX^L,
 \end{equation}
 where we use the notation
 $\bigwedge_{i=1}^{n}dX^i=dX^1\land\cdots\land dX^n$.

 In the terms of the incidence matrices
 a magnetic field is assumed in the form
\begin{equation}\label{maganz}
 F_b^{({\cal M})\;K_1\ldots K_{d+1}}=\frac{1}{\sqrt{-g}}v_be^{\alpha\phi}
 \epsilon^{K_1\ldots K_{d+1} \beta}\partial_{\beta}e^{\chi},
\end{equation}
 where we take $K_i$ such that $\Lambda_{bK_i}=1$.

 We will use Einstein equations in the form $R_{KL}=G_{KL}$,
 there $G_{KL}$ is related with the stress-energy tensor $T_{KL}$ as
 \begin{equation}
   G_{KL}=T_{KL}-\frac{g_{KL}}{D-2}T_P^{~~P}.
 \end{equation}
 For the considered ansatz the tensor $G$ is
 \begin{eqnarray}\label{tensorG}
  &G_{KL}=\frac{1}{2}\partial_K\phi\partial_L\phi&\nonumber\\&
         +\sum_{a=1}^{E}\frac{h_a^2}{2}
         e^{2F_L-2B+{\cal F}_a}
         \left(
           -\partial_K C_a\partial_L C_a
           -\eta_{KL}
           \left\{
             \Delta_{aL}-\frac{d}{D-2}
           \right\}
           (\partial C_a)^2
         \right)\nonumber&\\&
         +\sum_{b=1}^{M}\frac{v_b^2}{2}
         e^{2F_L-2B+{\cal F}_b}
         \left(
           -\partial_K \chi_b\partial_L \chi_b
           +\eta_{KL}
           \left\{
             \Lambda_{bL}-\frac{d}{D-2}
           \right\}
           (\partial \chi_b)^2
         \right),&
 \end{eqnarray}
 where
 \begin{eqnarray}
 &&{\cal F}_a=-\alpha\phi-2\sum\limits_{N=0}^{D-1}
                    \Delta_{aN}F_N+2C_a,\nonumber\\&&
 {\cal F}_b=\alpha\phi+2\sum\limits_{N=0}^{D-1}
                    \Lambda_{bN}F_N-4B+2\chi_b.
 \label{calF}
 \end{eqnarray}
 In order to guarantee the above form of the $G$-tensor
 (the absence of the $(ij)$-, $(\mu\nu)$-, $(\mu i)$-, $(\mu\alpha)$-
 and $(i\alpha)$-components,
 where $i\not=j$, $\mu\not=\nu$) we have to impose  additional
 restrictions on the incidence matrices. Namely, each two rows
 assumed to have differences in more then two columns.  Furthermore,
 the difference between each row of the electric incidence matrix and
 each row of the magnetic incidence matrix have to be in more then
 four columns.

 According to step 3 we assume the following ``no-force'' conditions
 \begin{eqnarray}
   \label{1.14}
   &&{\cal F}_a=0,
   \;a=1,\dots,E,\\
   \label{1.15}
   &&{\cal F}_b=0,
   \;b=1,\dots,M.
 \end{eqnarray}
 The LHS of these conditions are nothing but exponents which
 enter in the $G$ tensor. Using (\ref{1.15}) the magnetic field
 (\ref{maganz}) could be rewritten  in the following form
 \begin{equation}\label{1.2}
   F_b^{({\cal M})}=v_be^{-\chi_b(x)}\ast
 d\chi_b(x)\bigwedge_{\{i|\Lambda_{bi}=1\}} dy^i,
  \end{equation}
 where $\ast$ is a
 Hodge dual on the $x$-subspace.

 From Maxwell's equations and from Bianchi identity for $F$ under
 conditions (\ref{1.14}) and (\ref{1.15}) we get
 \begin{equation}\label{1.20}
   \Delta C_a=(\partial C_a)^2,\qquad
   \Delta \chi_b=(\partial\chi_b )^2,
 \end{equation}
 therefore, $H_{a}=e^{-C_{a}}$, $a=1,\dots,E$ and
 $U_{b}=e^{-\chi _{b}}$, $b=1,\dots,M$
 are harmonic functions.

 In order to solve the Einstein equations and the equation
 of motion for the dilaton we suppose
 \begin{eqnarray}
   \label{1.22}
   F_L&=&\sum_{a=1}^E h_a^2   C_a
         \left({1\over 2}\Delta _{aL}-u\right)
        -\sum_{b=1}^M v_b^2\chi_b
         \left({1\over 2}\Lambda_{bL}-u\right),\\
   \label{1.24}
   \phi&=&\frac{\alpha}{2}
          \left[
             \sum_{a=1}^{E}h_a^2    C_a
            -\sum_{b=1}^{M}v_a^2 \chi_b
          \right],
 \end{eqnarray}
 where
 \begin{equation}\label{u}
     u=\frac{d}{2(D-2)}.
 \end{equation}

 Substituting  (\ref{1.22}) and  (\ref{1.24}) in (\ref{gauge})
 one can check that equation (\ref{gauge}) is fulfilled.
 A substitution of (\ref{1.22}) and (\ref{1.24}) in the
 ``no force'' conditions under  assumption of the
 independence of $C_a$
 and $\chi_b$ gives three types of {\it characteristic}
 equations on the incidence matrices
 \begin{eqnarray}\label{7}
   &&\frac{\alpha^2}{2}-\frac{d^2}{D-2}
   +\sum_{L=0}^{D-1}\Delta_{aL}\Delta_{a'L}=0,
   ~~a\neq a',~~a,a'=1,\dots,E,\\
   \label{9}
   &&\frac{\alpha^2}{2}-\frac{d^2}{D-2}
   +\sum_{L=0}^{D-1}\Lambda_{bL}\Lambda_{b'L}-2=0,\;
   b\neq b',~~b,b'=1,\dots,M,\\
   \label{10}
   &&\frac{\alpha^2}{2}-\frac{d^2}{D-2}
   +\sum_{L=0}^{D-1}\Delta_{aL}\Lambda_{bL}=0,
   ~~a=1,\dots,E,~~b=1,\dots,M,
 \end{eqnarray}
 where $h_a^2=v_b^2=\sigma$,
 \begin{equation}\label{sigma}
   \sigma=\frac{1}{dt+\alpha^2/4},\qquad t=\frac{D-d-2}{2(D-2)}.
 \end{equation}
 Note that $\sum_{L=0}^{D-1}\Delta_{aL}\Delta_{a'L}$ is a
 number of common columns of $a$ and $a'$ branches in the
 incidence table. Characteristic equations admit solutions
 only for quantized values of  scalar coupling parameter
 that is in accordance with \cite{Berg,AV,IN}.

 Note that if the dilaton is absent, then the p-brane solutions are
 given by the  above formula with $\alpha =0$. Corresponding
 characteristic
 equations are very restrictive since they have to be solved
 for integers and they admit non-trivial solutions only for
 special $D$. In particular, they are
 $D=6,~10,~ 11,$ $18,~ 20,~ 26$.

 One can check that if the incidence matrices satisfy
 the characteristic equations then the Einstein equations
 hold. Therefore, the form of the metric which solves
 the theory is
 \begin{eqnarray}
  &&ds^2=\left(
               H_{1}H_{2}\cdots H_{E}\right)^{2u\sigma}
         \left(U_{1}U_{2}\cdots U_{M}\right)^{2t\sigma}
          \nonumber\\
  && \times\left\{
            \sum _{K,L=0}^{D-s-3}
            \left(
              \prod_{a}H_{a}^{\Delta_{aL}}
              \prod_{b}U_{b}^{1-\Lambda_{bL}}
            \right)^{-\sigma}
            \eta_{KL}dy^K dy^L
           +\sum_\gamma dx^\gamma dx^\gamma
          \right\}, \label{8'}
 \end{eqnarray}
 where $H_a$ and $U_b$  are harmonic functions,
 $u$, $t$ and $\sigma$ are given by (\ref{u}) and
 (\ref{sigma}).

 The explicit formula for the dilaton has the form
 \begin{equation}
  e^\phi=\left(
           \prod_{a}H_a^{-1}
           \prod_{b}U_b
         \right)^\frac{\alpha\sigma}{2}.
 \end{equation}

 The electric strength is
 \begin{equation}\label{ef}
   F^{(\cal E)}_{\alpha\mu_1^a\cdots\mu_{d}^a}=
         \mp\sqrt{\sigma} e^{\frac{1}{2}\alpha\phi}
          E^a_{\mu_1^a\cdots\mu_{d}^a}
                        \partial_\alpha \ln H_a,
 \end{equation}
 and the magnetic one is
 \begin{equation}\label{mf}
   F^{(\cal M)}_{\mu_0^b\cdots\mu_{d}^b}=
      \mp\sqrt{\sigma} e^{\frac{1}{2}\alpha\phi}
      E^b_{\mu_0^b\cdots\mu_{d}^b\alpha}
      g^{\alpha\beta}\partial_\beta \ln U_b,
 \end{equation}
 where $\mu^a\in\{L|\Delta_{aL}=1\}$,
 $\mu^b\in\{L|\Lambda_{bL}=1\}$,
 \begin{equation}
  E^a_{\mu_1^a\cdots\mu_{d}^a}=
  \sqrt{-\prod_{L=0}^{D-1}g_{LL}^{\Delta_{aL}}}
  \epsilon_{\mu_1^a\cdots\mu_{d}^a},
 \end{equation}
 \begin{equation}
    E^b_{\mu_0^b\cdots\mu_{d}^b\alpha}=
  \sqrt{\prod_{L=0}^{D-1}g_{LL}^{\Lambda_{bL}}}
  \epsilon_{\mu_0^b\cdots\mu_{d}^b\alpha}
 \end{equation}
  are forms of $d$ and $d+2$
 dimensional volume.

 The approach considered above could be obviously generalized to
 the actions with $k$ antisymmetric fields and several dilatons
 \begin{equation}\label{genact}
   I=\frac{1}{2\kappa ^{2}}
      \int d^{D}X\sqrt{-g}
      \left(
         R-\frac{1}{2}(\nabla\vec\phi)^2-
         \sum_{I=1}^{k}
         \frac{e^{-\vec\alpha^{(I)}\vec\phi}}{2(d_I+1)!}
         F^{(I)2}_{d_I+1}
      \right).
 \end{equation}
 In this case  we introduce $2k$ incidence matrices.
 Electric and magnetic fields are assumed in the form
 \begin{equation}\label{a1}
   A^{(I)}_a=h_a^{(I)}e^{C_a^{(I)}(x)}
           \bigwedge_{\{L|\Delta ^{(I)}_{aL}=1\}}dX^L,
 \end{equation}
 \begin{equation}\label{m1}
  F_b^{({I})\;K_1\ldots
 K_{d_I+1}}=\frac{1}{\sqrt{-g}}v_b^{(I)}e^{\vec\alpha^{(I)}\vec\phi}
 \epsilon^{K_1\ldots K_{d+1} \beta}\partial_{\beta}e^{\chi^{(I)}},
  \end{equation}
 where $I=1,\ldots,k$ and we take $K_i$ such that $\Lambda_{bK_i}^{(I)}=1$.
 Instead of (\ref{7}), (\ref{9}) and (\ref{10}) one has the following
 characteristic equations
 \begin{equation}\label{condee}
   (1-\delta_{IJ}\delta_{aa'})
   \left\{
     {\vec\alpha^{(I)}\vec\alpha^{(J)}\over 2}
     -{d_I d_J\over D-2}
     +\sum_{L=0}^{D-1}\Delta_{aL}^{(I)}\Delta_{a'L}^{(J)}
   \right\}=0,
 \end{equation}
 \begin{equation}\label{condmm}
   (1-\delta_{IJ}\delta_{bb'})
   \left\{
     {\vec\alpha^{(I)}\vec\alpha^{(J)}\over 2}
     -{d_Id_J\over D-2}
     +\sum_{L=0}^{D-1}\Lambda_{bL}^{(I)}\Lambda_{b'L}^{(J)}-2
     \right\}=0,
 \end{equation}
 \begin{equation}\label{condem}
   {\vec\alpha^{(I)}\vec\alpha^{(J)}\over 2}
   -{d_Id_J\over D-2}
   +\sum_{L=0}^{D-1}\Delta_{aL}^{(I)}\Lambda_{b'L}^{(J)}=0.
 \end{equation}
 The constants $h_a^{(I)}$ and $v_b^{(I)}$ are given by
 \begin{equation}\label{3.4}
   {h_a^{(I)}}^2={v_b^{(I)}}^2=\sigma^{(I)},
   \qquad\mbox{where}\qquad
   \sigma^{(I)}=\frac{1}{t^{(I)}d_I+\frac{1}{4}\left.\vec\alpha^{(I)}\right.^2},
\qquad t^{(I)}=\frac{D-2-d_I}{2(D-2)}.
\end{equation}
 Let us present the metric in
the form where the overall conformal factor
which multiplies the transverse $x$-part is separated
  \begin{eqnarray}\label{genmetr}
   &ds^2=\prod_{I=1}^k
        \left(
          H_{1}^{(I)}H_{2}^{(I)}\cdots H_{E_I}^{(I)}
        \right)^{2u^{(I)}
        \sigma^{(I)}}
        \left(
          U_{1}^{(I)}U_{2}^{(I)}\cdots U_{M_I}^{(I)}
        \right)^{2t^{(I)}
        \sigma^{(I)}}&\nonumber\\
  &\times\left\{
     \sum_{L=0}^{D-s-3}
     \prod_{I=1}^k
     \left(
       \prod _{a}{H_{a}^{(I)}}^{\Delta_{aL}^{(I)}}
       \prod _{b}{U_{b}^{(I)}}^{1-\Lambda_{bL}^{(I)}}
     \right)^{-\sigma^{(I)}}\eta_{KL}
     dy^K dy^L
     +\sum _{\gamma}dx^{\gamma}dx^{\gamma}
   \right\},&
\end{eqnarray}
  where
 \begin{equation}\label{t-u}
    u^{(I)}=\frac{    d_I}{2(D-2)}.
 \end{equation}

 Under assumption $s>0$ we take the harmonic functions
 in the form
 \begin{equation}
 \label{sb0}
  H_a^{(I)}=1+\sum_{c}\frac{Q_{ac}^{(I)}}{|x-x_{ac}|^s},\qquad
  U_b^{(I)}=1+\sum_{c}\frac{P_{bc}^{(I)}}{|x-x_{bc}|^s}.
 \end{equation}

 The algebraic method \cite{AV,AVV,IN,AVVV} can be also used for
 finding solutions with depending harmonic functions.

%%%+++++++++++++++++++++++++++++++++++++++++++++++++++%%%
\section{ADM mass and area of horizon}
 The representation (\ref{genmetr}) is convenient for calculating
 the ADM mass and the entropy. Let the harmonic functions
 have the form (\ref{sb0}) and all these functions have
 the same centers ($Q^{(I)}_{ac},P^{(I)}_{bc}>0$ for
 all $a,b,c,I$).

 The ADM mass has the form
 \begin{equation}
   M=\frac{L^{d-1}\omega_{s+1} s}{2\kappa^2}
                 \sum_I \sigma^{(I)}
     \left(\sum_{ac} Q_{ac}^{(I)}+\sum_{bc} P_{bc}^{(I)}\right),
 \end{equation}
 where $\omega_{s+1}$ is a volume of the $s+1$-dimensional sphere,
 \begin{equation}
  \omega_{s+1}=\frac{2\pi^\frac{s+2}{2}}
               {\Gamma\left(\frac{s+2}{2}\right)}
 \end{equation}
 and $L$ is a period of all $y_i$, $i=1,\dots,D-s-3$.

 Under the condition
 \begin{equation}
 \label{Ac}
	 \frac{s}{2}\sum_I \sigma^{(I)}(E_I+M_I)=s+1
 \end{equation}
 the area of horizon has the form
 \begin{equation}
	 {\cal A}_{D-2}=\omega_{s+1}L^{D-s-3}
                  \sum_c\prod_I
                  \left(
                    \prod_a Q^{(I)}_{ac}
                 \prod_b P^{(I)}_{bc}
         \right)^\frac{\sigma^{(I)}}{2},
 \end{equation}

 One can also get non-trivial entropy if $q\geq 2$ and
 instead of (\ref{Ac})
 the following condition is assumed
 \begin{equation}
	 \frac{s}{2}\sum_I \sigma^{(I)}(E_I+M_I)=\frac{s}{2}+1.
 \end{equation}
 To get non-trivial entropy we make a boost
 \begin{equation}
   -dy_0^2+dy_1^2\;\longrightarrow\;dudv+K(x)du^2,
 \end{equation}
 where
 \begin{equation}
   u=y_1+y_0,\;
   v=y_1-y_0,\;
   K(x)=\sum_c\frac{Q_c}{|x-x_c|^s},\;
   x_c=x_{ac}.
 \end{equation}
 In this case the area of gorizon is given by
 \begin{equation}
	 {\cal A}_{D-2}=\omega_{s+1}L^{D-s-3}
                  \sum_c Q_c^\frac{1}{2} \prod_I
                  \left(
                \prod_a Q^{(I)}_{ac}
            \prod_b P^{(I)}_{bc}
        \right)^\frac{\sigma^{(I)}}{2}.
 \end{equation}

\section{$S$-duality}
 In order to demonstrate $S$-duality let us consider a new action, which
 could be obtained from the action (\ref{genact}) by replacing
 an antisymmetric field $F^{(I)}_{d_I}$ by another
 field $F^{(I)}_{\tilde d_{I}}$,
 \begin{equation}
 \label{l1}
   \tilde d_I=D-2-d_I,
 \end{equation}
 and changing
 the signs of the corresponding dilaton coupling constants on the
 opposite ones:
 \begin{equation}
 \label{l2}
   \tilde{\vec\alpha}^{(I)}=-\vec\alpha^{(I)}.
 \end{equation}
 $S$-duality
 transforms the solutions of the theory  (\ref{genact}) into the
 solutions of the theory with a new action. The corresponding transformations
 of the incidence matrices are
 \begin{eqnarray}
 \label{trF}
  \Delta_{aL}^{(I)}&\rightarrow&\tilde\Delta_{bL}^{(I)}=1-\Lambda_{bL}^{(I)},\\
 \label{trL}
  \Lambda_{bL}^{(I)}&\rightarrow&\tilde\Lambda_{aL}^{(I)}=1-\Delta_{aL}^{(I)}.
 \end{eqnarray}
 One can check that the new incidence matrices also satisfy the
 characteristic equations.

 Also one can perform $S$-duality transformation
 (\ref{trF})--(\ref{trL})  only for  some branches of the fields.
 In this case the dual theory may have more fields in comparison
 with the initial one.

%%%---------------------------------------------------%%%

$S$-duality becomes evident if we present
our results in the p-brane terminology. In order to
consider the electric and magnetic p-branes together we introduce
 the {\it brane incidence matrix} $\Upsilon_{PL}$, where
 $L=0,\ldots,D-1;\;P=1,\ldots,B$, $B$ is a whole number of p-branes.
The brane incidence matrix is constructed in the following way
\begin{equation}
  \Upsilon_{PL}=
    \left(
      \begin{array}{c}
        \Delta_{aL}^{(1)}\\
        \Delta_{aL}^{(2)}\\
        \vdots\\
        \Delta_{aL}^{(k)}\\
        1-\Lambda_{bL}^{(1)}\\
        1-\Lambda_{bL}^{(2)}\\
        \vdots\\1-\Lambda_{bL}^{(k)}
      \end{array}
    \right).
\end{equation}
The entries of the brane incidence matrix are equal to 1 or 0. The
rows of this matrix correspond to p-branes. Electrically charged p-branes
occupy the upper rows of the matrix, magnetically charged p-branes
occupy the down rows. The columns of the matrix $\Upsilon_{PL}$
correspond to  the space-time indices (similar to $\Delta$
and $\Lambda$).  This matrix could be represented as the brane
incidence table (see example (\ref{brtab2})). We denote by "$\times$"
electrically charged p-branes and by "$\ast$" magnetically charged
ones. For $\Upsilon_{PL}$ we have only one characteristic equation
instead of three ones. Namely, the equations (\ref{condee}),
(\ref{condmm}) and (\ref{condem})  in the terms of  $\Upsilon_{PL}$
could be rewritten in the following form
\begin{equation}\label{condbr}
   \varsigma_R\varsigma_{R'}
   {\vec\alpha^{(R)}\vec\alpha^{(R')}\over 2}
  -{d_R d_{R'}\over D-2}
  +\sum_{L=0}^{D-1}\Upsilon_{RL}\Upsilon_{R'L}=0,\;R\not=R',
\end{equation}
where $\varsigma_R$ is $-1$ for the electric p-branes and $+1$ for the
magnetic ones, $\vec\alpha^{(R)}$ are dilatons coupling constants
connected with the $p_R$-brane, $d_R=p_R+1$.  This unified form of
the characteristic equations is a manifestation of the $S$-duality.

\section{ Harmonic function rule}
Our solution (\ref{genmetr}) has a very simple structure.
This becomes obvious if one rewrites the metric in the
following form:

\begin{equation}\label{hfrF}
   g_{KK}=\prod_I\prod_a g_{KK}^{(I)a} \prod_b g_{KK}^{(I)b},
\end{equation}
where
\begin{eqnarray}
   g_{KK}^{(I)a}&=&\left(H^{(I)\Delta_{aK}}_a\right)^{\tau^{(I)}}
             \left(H^{(I)(1-\Delta_{aK})}_a\right)^{\rho^{(I)}},\\
\label{hfrL}
   g_{KK}^{(I)b}&=&\left(U^{(I)\Lambda_{bK}}_b\right)^{-\tau^{(I)}}
             \left(U^{(I)(1-\Lambda_{bK})}_b\right)^{-\rho^{(I)}}.
\end{eqnarray}

According to (\ref{3.4}) and (\ref{t-u}) the exponents are given by:
\begin{equation}
\tau^{(I)} \equiv -2t^{(I)}\sigma^{(I)}=-{4(D-2-d_I)\over \Delta^{(I)}(D-2)},
\quad
\rho^{(I)} \equiv  2u^{(I)}\sigma^{(I)}={4d_I\over\Delta^{(I)}(D-2)},
\end{equation}
where $\Delta^{(I)}$ is a generalization of (\ref{Dlt})
\begin{equation}
\Delta^{(I)}=\vec\alpha^{(I)2}+{2d_I(D-2-d_I)\over D-2}.
\end{equation}

For given incidence
matrices and values of $\tau^{(I)}$ and $\rho^{(I)}$
(\ref{hfrF})--(\ref{hfrL}) gives the
following rule for constructing a metric.  For each space-time
direction the coefficient in the metric is a product of functions $H_a$ and
$U_b$ in an appropriate power.  Namely, we put
$\left.H_n^{(I)}\right.^{\tau^{(I)}}$
($\left.U_n^{(I)}\right.^{-\rho^{(I)}}$)
if the corresponding direction belongs to the $n$-th
$(d-1)-$ electric ($(D-d-3)-$ magnetic) brane,
and we put $\left.H_n^{(I)}\right.^{\rho^{(I)}}$
($\left.U_n^{(I)}\right.^{-\tau^{(I)}}$)
if the corresponding direction is transverse to
$(d-1)-$ electric ($(D-d-3)-$ magnetic) brane. Note that $\tau^{(I)}$ and
$\rho^{(I)}$  are the same as in the corresponding single brane.
Certainly, one has to assume that the incidence matrices satisfy
the characteristic equations (\ref{condee})--(\ref{condem}).

There is another point of view, based on  the metric representation
 (\ref{genmetr}).  The overall conformal factor
in (\ref{genmetr}) can be rewritten as
\begin{eqnarray}
&\prod_{I=1}^k \left(
          H_{1}^{(I)}H_{2}^{(I)}\cdots H_{E_I}^{(I)}
        \right)^{2u^{(I)}
        \sigma^{(I)}}
        \left(
          U_{1}^{(I)}U_{2}^{(I)}\cdots U_{M_I}^{(I)}
        \right)^{2t^{(I)}
        \sigma^{(I)}}&\nonumber\\
&=\prod_{I=1}^k
        \left(
          H_{1}^{(I)}H_{2}^{(I)}\cdots H_{E_I}^{(I)}
        \right)^{\frac{d_{I}}{D-2}
        \sigma^{(I)}}
        \left(
          U_{1}^{(I)}U_{2}^{(I)}\cdots U_{M_I}^{(I)}
        \right)^{\frac{D-d_{I}-2}{D-2}
        \sigma^{(I)}}.&
  \end{eqnarray}
Since the magnetic p-brane is connected with the $(D-p-3)$-form, the
conformal factor could be unified
\begin{equation}
\prod_{R=1}^{B} H_R^{\frac{p_R+1}{D-2}\sigma^{(R)}},
\end{equation}
where we have a product of all harmonic functions in an appropriate power
which is independent on the p-brane charge and is determined by the
p-brane dimension. For each space-time dimension the coefficient in
the metric (in brackets) is also a product of harmonic functions in
power $\sigma^{(R)}$. Namely, we put $H_R^{-\sigma^{(R)}}$ if the
corresponding direction belongs to the $R$-th brane. So the form of
the metric doesn't depend on the type of p-brane charges and could be
constructed in terms of the brane incidence matrix (which describes p-brane
intersections only)
\begin{equation}\label{Upsmetr}
 ds^2=\prod_{R=1}^{B}
 H_R^{\frac{p_R+1}{D-2}\sigma^{(R)}}
 \left\{
   \sum_{L=0}^{D-s-3}
   \left(
     \prod_{R=1}^B
     H_R^{\Upsilon_{RL}}
   \right)^{-\sigma^{(R)}}\eta_{KL}
   dy^Kdy^L
   +\sum _{\gamma}
   dx^{\gamma}dx^{\gamma}
 \right\}.
\end{equation}

 Therefore
the formula (\ref{Upsmetr}) gives a $D$-dimensional generalization of the
``harmonic function rule'' found before for $D=10,\;11$ ~\cite{Ts1}.

\section{$T$-duality}
Let us
consider generalized $T$-duality transformations.  $T$-duality transforms
solutions for the action (\ref{genact}) with one set of fields into
solutions of the action (\ref{genact}) with another set of fields.
 We perform
$T$-duality transformation along the direction
corresponding to $y_{i_0}$ coordinate, $q\le i_0 \le D-s-3$.
T-duality acts on the brane incidence
matrix $\Upsilon_{RL}$ as follows. We select the $i_0$-th column,
 change 1 into 0 and vice versa
and  obtain a new brane incidence matrix. This matrix satisfies the
characteristic equation if we simultaneously  change dilaton coupling
constants.  More precisely, new dilaton coupling constants $\vec\beta_R$  are
connected with the  old ones  $\vec\alpha_R$  in the following way
\begin{eqnarray}
&\frac{\vec\beta_R\vec\beta_{R'}}{2}
=\frac{\vec\alpha_R\vec\alpha_{R'}}{2}-1+
\Upsilon_{Ri_0}+\Upsilon_{R'i_0}+\frac{1}{D-2}\left[(1-2\Upsilon_{Ri_0})
(1-2\Upsilon_{R'i_0})\right.&\nonumber\\
& \label{newalph}
\left.+(1-2\Upsilon_{Ri_0})d_{R'}+(1-2\Upsilon_{R'i_0})d_{R}
\right].&
 \end{eqnarray}
 In particular cases these relations give rather restrictive conditions
  on the initial theory parameters. For example, let us consider
  the case of one dilaton and one antisymmetric field
  and let us deal with the electrically charged p-branes.
   We make
  the mapping \begin{equation} \label{Tdual}
  I(D,d,\alpha)\longmapsto I(D,d-1,d+1,\alpha_1,\alpha_2)
   \end{equation}
 and call it as a generalized  $T$-duality transformation.
 $T$-duality ~\cite{HT,W,PT,JHS,Sen,BO}
in the case of IIA superstring transforms solutions
with non-zero 2-form into solutions with non-zero 1- and 3-forms
 ~\cite{BergRoo}. After performing $T$-duality we have a new brane incidence
 matrix $\Upsilon_{RL}'$ which is effectively a composition of the brane
incidence matrices for $(d-2)$- and $d$-branes.  In the case $E'>1,\;E''>1$
 the  characteristic equations   (\ref{newalph})
  are consistent only if
  \begin{equation}\label{alphcond}
 \alpha^2={(2d-D+2)^2\over 2(D-2)}
\end{equation}
 (for $E'=1$ or $E''=1$ the characteristic equations are less restrictive).
 Only under condition (\ref{alphcond})
 $T$-duality in the form (\ref{Tdual}) takes place. Using (\ref{8'}) and
 (\ref{genmetr})  one can check that the values of $\sigma$ which specify
 the relative transverse components of
the metrics corresponding to the theories related by $T$-duality are the
same.

\section{The case of arbitrary signature}
\label{asig}
p-brane solutions also exist in the case of an arbitrary
space-time signature:  $\eta={\rm diag}(\pm 1,\dots,\pm 1)$.
In this case one deals with a modified action
\begin{equation}
 I=\frac{1}{2\kappa ^{2}}
   \int d^{D}X\sqrt{|g|}
   \left(
     R-\frac{1}{2}(\nabla\vec\phi)^2-
     \sum_{I=1}^{k}
     \frac{s_I e^{-\vec\alpha^{(I)}\vec\phi}}{2(d_I+1)!}
     F^{(I)2}_{d_I+1}
   \right),
\end{equation}
 where $s_I=\pm 1$. Instead of old restrictions for the
 incidence matrices one has new ones
 \begin{eqnarray}
    &s_I\prod\limits_{L=0}^{D-1}(\eta_{LL})^{\Delta_{aL}^{(I)}}=-1\qquad
    &\mbox{instead of}\qquad
    \Delta_{a0}^{(I)}=1,\\
    &s_I\prod\limits_{L=0}^{D-1}(\eta_{LL})^{\Lambda_{bL}^{(I)}}=+1\qquad
    &\mbox{instead of}\qquad
    \Lambda_{b0}^{(I)}=0.
 \end{eqnarray}
 In the unified notation we may rewrite this restrictions as
 \begin{equation}
 \label{s_i}
   s_I\prod\limits_{L=0}^{D-1}(\eta_{LL})^{\Delta_{RL}}=\varsigma_R,
 \end{equation}
 where $\Delta_{RL}$ is the {\it matter incidence matrix}, which contain
 rows corresponding to both types of p-branes, electric and magnetic.
 The characteristic equations remain the same.

 Under the $S$-duality transformation
 in addition to (\ref{l1}),(\ref{l2}) one has also to perform a change
  \begin{equation}
  s_I\rightarrow \tilde s_I=-{\rm det}(\eta_{KL}) s_I.
 \end{equation}

\section{Supersymmetry in 11D supergravity}
Let us recall that some p-brane solutions were obtained using requirements
of supersymmetry. Known supersymmetric p-brane solutions admit an existence
of Killing spinors. In this Section we are going to examine relations
between the problem of finding Killing spinors and our scheme of finding
solutions taking as example 11D supergravity.
  The bosonic sector of $D=11$ supergravity consists
 of a metric and a three-form potential.
  Killing spinors $\varepsilon$ satisfy the following equations
 \begin{eqnarray}
                \tilde D_L\varepsilon&=&0,\\
                \tilde D_L&=&\partial_L+\frac{1}{4}\omega_L^{~~AB}\Gamma_{AB}
                                -\frac{1}{288}(\Gamma^{PQRS}_{~~~~~~L}
         +8\Gamma^{PQR}\delta^S_L)
				F_{PQRS},
 \end{eqnarray}
         where $P,Q,R,S,L$ are 11D world indices. $\Gamma_A$ are
	 the 11D Dirac matrices
 \begin{eqnarray}
   \{\Gamma_A,\Gamma_B\}=2\eta_{AB},\\
   \Gamma_{AB\cdots C}=\Gamma_{[A}\Gamma_{B\cdots}\Gamma_{C]}.
 \end{eqnarray}
 Indices $A,B,C$ are 11D vielbein indices,
 related with the orthonormal vielbein.

 If we assume the ansatz for the metric (\ref{3}) and
 for the matter field (\ref{1.1}), (\ref{maganz}), then
 in the Fock--De~Donder gauge the covariant derivative is
 \begin{eqnarray}\label{der}
   &&\tilde D_L=\partial_L-\frac{1}{2}\Gamma_L^{~~K}\partial_K F_L
          +\sum_a
          \frac{h_a}{6}e^{{\cal F}_a/2}
          \partial_L C_a \Gamma(a)-
          \sum_b
          \frac{v_b}{12}e^{{\cal F}_b/2}
          \partial_L \chi_b \Gamma(b)\\
          &&+\Gamma_L^{~~K}
          \sum_a \frac{h_a}{2}
          \left\{\Delta_{aL}-\frac{1}{3}\right\}
          e^{{\cal F}_a/2}
          \partial_K C_a \frac{\Gamma(a)}{2}
          +\Gamma_L^{~~K}
          \sum_b\frac{v_b}{2}
          \left\{\Lambda_{bL}-\frac{1}{3}\right\}
          e^{{\cal F}_b/2}
          \partial_K \chi_b \frac{\Gamma(b)}{2},\nonumber
 \end{eqnarray}
 where
 \begin{eqnarray}
   \Gamma(a)=\frac{1}{3!}E^{\mu^a_1\mu^a_2\mu^a_3}
             \Gamma_{\mu^a_1\mu^a_2\mu^a_3}=
             \prod_{\{A|\Delta_{aA}=1\}} \Gamma_A,\\
   \Gamma(b)=\frac{1}{5!}E^{\mu^b_1\mu^b_2\mu^b_3\mu^b_4\mu^b_5}
             \Gamma_{\mu^b_1\mu^b_2\mu^b_3\mu^b_4\mu^b_5}=
             \prod_{\{A|\Lambda_{bA}=1\}}\Gamma_A.
 \end{eqnarray}
 Here notations are the same as in (\ref{ef}), (\ref{mf}).
 For definition of ${\cal F}_a$ and ${\cal F}_b$
 see (\ref{calF}).

 To analyze the consequences of the supersymmetry
 it is convenient to rewrite the covariant
 derivative as
 \begin{eqnarray}\label{derU}
   \tilde D_L&=&\partial_L-\frac{1}{2}\Gamma_L^{~~K}\partial_K F_L
              + \sum_R
                \frac{1-3\varsigma_R}{24}h_R e^{{\cal F}_R/2}
                \partial_L C_R \Gamma(R)\\
             \nonumber
             &+&\Gamma_L^{~~K}
                \sum_R \frac{h_R}{2}
                \left\{\Delta_{RL}-\frac{1}{3}\right\}
                e^{{\cal F}_R/2}
                \partial_K C_R \frac{\Gamma(R)}{2}.
 \end{eqnarray}
 The notations in (\ref{derU}) are obvious
 $$
   C_R=\left\{
         \begin{array}{ll}
           C_a,   &R=a\\
           \chi_b,&R=b
         \end{array}
       \right. ,
   \qquad
   h_R=\left\{
         \begin{array}{ll}
           h_a,&R=a\\
           v_b,&R=b
         \end{array}
       \right. .
 $$
 Definition of the matrix $\Delta_{RL}$ one can find
 after equation (\ref{s_i}).
 Note that in the expression (\ref{derU}) for the
 covariant derivative we do not assume that the
 metric solves the Einstein equations.

 To construct the Killing spinor it is natural to
 assume that all ${\cal F}_R$ vanish. These conditions
 coincide with ``no force'' conditions
 (\ref{1.14}), (\ref{1.15}).

 Now we can see that if we assume that there exist
 spinors $\varepsilon_0$ such that
 \begin{equation}
 \label{eq80}
    \Gamma_L^{~~K}
    \left[
      -\frac{1}{2}\partial_K F_L
      +\sum_R
       \frac{h_R}{2}
       \left\{\Delta_{RL}-\frac{1}{3}\right\}
       \partial_K C_R
       \frac{\Gamma(R)}{2}
    \right]
    \varepsilon_0=0,
 \end{equation}
 then the Killing spinors can be found in the form
 \begin{equation}
   \varepsilon=\varepsilon_0 f(x),
 \end{equation}
 where a scalar function $f(x)$ solves the following
 equations
 \begin{equation}
   \left[
     \partial_L
     +\sum_R
     \frac{1-3\varsigma_R}{24}
     h_R
     \partial_L C_R
     \Gamma(R)
   \right]
   \varepsilon_0 f(x)=0.
 \end{equation}
 In order to solve the equation (\ref{eq80}),
 let us assume that $F_L$ are linear combinations
 of the functions $C_R$ (a ``relax'' harmonic
 superposition rule),
 for simplification of our calculation we shall write
 $F_L=-1/2\sum l_{RL}h_R
          \{\Delta_{RL}-1/3\} C_R$
 (we assume that $h_R\not=0$).
 Under this assumption and taking into account
 the independence of functions $C_R$ we get
 \begin{equation}\label{ttt}
    [l_{RL}+\Gamma(R)]\varepsilon_0=0.
 \end{equation}
 Therefore constants $l_{RL}$ do not depend on subscript $L$,
 i.e.
 \begin{equation}\label{tt}
    [l_{R}+\Gamma(R)]\varepsilon_0=0.
 \end{equation}
 This equation admits non-trivial solutions
 only for $|l_R|=1$. Note that substituting the relax
 harmonicity conditions in ``no force'' conditions one can
 find a relation $l_R h_R=\varsigma_R$ and the characteristic
 equation. Since $|l_R|=1$, we get $h_R^2=1$,
 $l_R=\varsigma_R h_R$. So under ``no force''
 conditions the relax harmonic superposition
 rule coincides with the harmonic superposition rule and
 one can say that the requirement of the supersymmetry supports the
 harmonic superposition.

 To guarantee that a configuration obtained
 as a result of intersecting of single branes  is supersymmetric
 one has to study a compatibility conditions
 of all requirements (\ref{tt}).
 In order to analyse these requirements
 let us introduce $S$ as a special set of signs
  \begin{equation}
   S=\{\pm,\dots,\pm\},~~\mbox{or}~~~S(R)=\pm.
 \end{equation}
 We search a set $S$ which admits an existence of $\varepsilon_0$
 such that
 \begin{equation}
 \label{susy-r}
   P_R^{S(R)}\varepsilon_0=0,
 \end{equation}
 for all $R$. Here $P_R^{S(R)}$ is a projector
 \begin{eqnarray}
   P_R^{S(R)}=\frac{1+S(R)\Gamma(R)}{2}.
 \end{eqnarray}
 We will demonstrate, that one can find an
 appropriate $S$ and $\varepsilon_0$, if
 the incidence matrices satisfy the characteristic equations.
 In this case of 11D supergravity they lead to the restrictions
 on the incidence matrix
 \begin{equation}
   \sum_{N=0}^{D-1}\Delta_{RN}\Delta_{R'N}=1,
   ~~~~\mbox{or}~~~~
   \sum_{N=0}^{D-1}\Delta_{RN}\Delta_{R'N}=3,
   ~~~~~R\not=R'.
 \end{equation}
 If the incidence matrix satisfies these conditions
 one can check that
 \begin{equation}
 \label{commG}
   [\Gamma(R),\Gamma(R')]=0.
 \end{equation}

 Using (\ref{commG}) one can introduce
 new projection operators
 \begin{equation}
   P^{S}=\prod_{R}P_R^{S(R)}
 \end{equation}
 with the following properties:
 \begin{eqnarray}
   P^{S}P^{S'}&=&\delta_{SS'}P^{S},\\
 \label{sum-P}
   \sum_S P^{S}&=&1.
 \end{eqnarray}
 Acting by $\sum_S P^{S}$ on an arbitrary
 spinor $\varepsilon_i\not=0$ and using (\ref{sum-P}),
 one can obtain the existence of the set $S_0$,
 such that $P^{S_0}\varepsilon_i=\varepsilon_0\not=0$.
 Moreover for every $R$ $P_R^{-S_0(R)}\varepsilon_0=0$,
 comparing this result with the equation (\ref{susy-r})
 one can conclude, that $-\varsigma_R S_0(R)$
 is supersymmetric signs set for $h_R$.

 To summarize, using a simple algebraic method
 we  have  constructed the general D-dimensional intersecting
 p-brane solutions which satisfy the
 harmonic function superposition rule and possess $S$- and $T$-dualities.
 The intersections of p-branes are controlled by
 the characteristic equations.
 These equations have solutions only for quantized values of  scalar coupling
 parameters. Some of these solutions in the cases of D=11 and D=10
 provide the metrics with regular horizons and non-zero entropy.

\section*{Acknowledgments}
We are grateful to I.~V.~Volovich for useful discussions.
This work is partially supported by the RFFI grants 96-01-00608
(I.A. and O.R.) and 96-01-00312 (M.I.).

\end{document}